\documentclass[10pt,aps,twocolumn,floatfix,prl,twoside,tightenlines,nofootinbib,showpacs]{revtex4}
\usepackage[sort&compress]{natbib}
\usepackage{graphicx,amsmath,amssymb,amsopn,bm}
\usepackage{color}

\renewcommand{\d}{\text{d}}
\newcommand{\im}{\text{Im}}
\newcommand{\ii}{\text{i}}
\newcommand{\erw}[1]{\ensuremath { %
    \left \langle {#1} \right \rangle}}
\newcommand{\op}[1]{\ensuremath{\bm{\mathrm{#1}}}}

\begin{document}

\title{Comprehensive Interpretation of Thermal Dileptons at the SPS} 

\author{Hendrik van Hees and Ralf Rapp} 
\affiliation{Cyclotron Institute and Physics Department, Texas A{\&}M
  University, College Station, Texas 77843-3366, USA}

\date{\today}

\begin{abstract}
  Employing thermal dilepton rates based on medium-modified
  electromagnetic correlation functions we show that recent dimuon
  spectra of the NA60 collaboration in central In-In collisions at the
  CERN-SPS can be understood in terms of radiation from a hot and dense
  hadronic medium.  Earlier calculated $\rho$-meson spectral functions,
  as following from hadronic many-body theory, provide an accurate
  description of the data up to dimuon invariant masses of about
  $M$$\simeq$0.9~GeV, with good sensitivity to details of the predicted
  $\rho$-meson line shape. This, in particular, identifies
  baryon-induced effects as the prevalent ones.  We show that a reliable
  description of the $\rho$ contribution opens the possibility to study
  further medium effects: at higher masses ($M$$\simeq$0.9-1.5~GeV)
  4-pion type annihilation is required to account for the experimentally
  observed excess (possibly augmented by effects of ``chiral mixing''),
  while predictions for thermal emission from modified $\omega$ and
  $\phi$ line shapes may be tested in the future.


\end{abstract}

\maketitle

\textit{Introduction.}  Electromagnetic probes occupy a special role in
the study of strongly interacting matter produced in energetic
collisions of heavy nuclei: once produced, photons and dileptons leave
the reaction zone essentially undistorted, carrying direct information
from the hot and dense medium to the detectors. While the ultimate goal
in analyzing pertinent experiments is to infer signatures of QCD phase
transitions (chiral symmetry restoration and/or deconfinement), the more
imminent (and relatively easier) objective is to extract medium
modifications of the electromagnetic (e.m.)  spectral
function~\cite{Rapp:1999ej,Brown:2003ee}. Recent data from the NA60
collaboration~\cite{Damjanovic:2005ni} on dimuon invariant-mass spectra
in In-In collisions at the CERN Super-Proton-Synchrotron (SPS) have
raised the experimental precision to an unprecedented level, posing
serious challenges to theoretical models. After subtraction of late
hadronic decay sources (``cocktail''), the excess dimuon radiation
exhibits features of a (broadened) peak around the free $\rho$-mass
($m_\rho$$\simeq$0.77~GeV) with a substantial enhancement at both lower
and higher invariant masses. Theoretical
predictions~\cite{Rapp:1999us,Rapp:2004zh} based on an expanding
fireball have confirmed that a broadened $\rho$-meson, as following from
hadronic many-body theory, is in line with the NA60 dimouns, while a
dropping $\rho$-mass~\cite{Cassing:1995zv,Li:1996db}, characterized by a
spectrum centered around a mass of 0.4-0.5~GeV, is inconsistent with the
data\footnote{A reinterpretation of the dropping-mass
  scenario~\cite{Brown:2003ee} within a Hidden Local Symmetry approach
  for vector mesons~\cite{Harada:2003jx} has not been confronted with
  dilepton data yet. However, any dropping-mass scenario will face the
  challenge of accounting for the large portion of strength seen in the
  NA60 data around the free $\rho$-mass.}. Both scenarios were
consistent with earlier dilepton data by the CERES
collaboration~\cite{Agakichiev:2005ai}.

The objective of the present letter is twofold: First, we improve on
earlier (shape-based) comparisons of our predictions to data based on
in-medium $\rho$ spectral functions. With a slight modification of the
expanding fireball model, we quantitatively reproduce (shape and
absolute magnitude of) the low-mass NA60 data in central In-In
collisions.
The spectral shape turns out to be sensitive to properties of the $\rho$
in some detail. Second, having determined in this way the contributions
of the 2-$\pi$ piece to the in-medium e.m.~correlator, we investigate
remaining enhancements in the $\mu^+\mu^-$ spectrum. Above
$M_{\mu\mu}$$\simeq$0.9~GeV, using empirical fits to the
isovector-vector ($V$) and -axialvector ($A$) spectra in vacuum, a
calculation with the free emission rate underestimates the data. When
including medium effects due to chiral $V$-$A$ mixing~\cite{Dey:1990ba},
with an assumed critical temperature of $T_c$=175~MeV (consistent with
the fireball), the description improves. While more precise conclusions
have to await more accurate data and calculations, our estimates
indicate that dilepton radiation from heavy-ion collisions at the SPS
emanates from matter close to the expected QCD phase boundary. Finally,
we address medium effects on the narrow vector mesons $\omega$ and
$\phi$.

\textit{Thermal Dileptons and Medium Effects.} The differential rate for
thermal lepton-pair production per unit 4-volume and 4-momentum can be
expressed as~\cite{MT84,wel90,gale-kap90}
\begin{equation}
\label{rate}
\frac{\d N_{ll}}{\d^4 x \d^4 q}
=-\frac{\alpha^2}{3 \pi^3} \frac{L(M^2)}{M^2}~\im
\Pi_{{\rm em},\mu}^{\mu}(M,q)~f^B(q_0;T)
\end{equation}
in terms of the retarded e.m.~current-current correlator
\begin{equation}
\label{ret-se}
\Pi^{\mu \nu}_{\rm em}(q)=\ii \int \d^4 x \mathrm{e}^{\ii q \cdot x}
\Theta(x^0) \erw{[\op{J}^{\mu}_{\rm em} (x),\op{J}^{\nu}_{\rm em}(0)]} 
\ , 
\end{equation}
a final-state lepton phase-space factor, $L(M)$,
and the thermal Bose function, $f^B(q_0;T)$. $\alpha$=$e^2/(4
\pi)$=1/137 denotes the fine structure constant and
$M^2$=$q_0^2$$-$$q^2$ the dilepton invariant mass with energy $q_0$ and
3-momentum $q$. The central objective in the following is the evaluation
of medium modifications of the e.m.~spectral function.

In the low-mass region (LMR, $M$$\le$1~GeV), the free e.m.~correlator is
saturated by the light vector mesons $\rho$, $\omega$ and $\phi$.  Our
main focus is on the $\rho$-meson (which dominates the dilepton yield in
the LMR), but we also investigate radiation from (and medium effects on)
the $\omega$ and $\phi$.  Although their contribution is suppressed
relative to the $\rho$,
it is an inevitable consequence of our assumption of the formation
of a thermalized medium.

Medium modifications of the $\rho$-meson are implemented using hadronic
many-body theory~\cite{Rapp:1999us}, accounting for interactions in hot
hadronic matter via (i) a dressing of its pion cloud with baryon-hole
excitations and thermal Bose factors, (ii) direct resonances on
surrounding mesons ($\pi$, $K$, $\rho$) and baryons (nucleons,
$\Delta$'s, hyperons, etc.). The effective interaction vertices
(coupling constants and form factors) have been carefully constrained by
a combination of hadronic and radiative decay branchings, by
photoabsorption on nucleons and nuclei, and by $\pi N$$\to$$\rho N$
scattering. The resulting $\rho$ spectral functions in cold nuclear
matter comply with constraints from QCD sum rules~\cite{Leupold:1997dg};
they have successfully been employed to dilepton spectra at full SPS
energy~\cite{Agakichiev:2005ai}, and to predict an even larger
enhancement at lower SPS energies~\cite{Rapp:2002mm,Adamova:2002kf}.
When averaged over a typical space-time evolution, the in-medium $\rho$
width at SPS amounts to $\Gamma_\rho^{\rm med}$$\simeq$$350$~MeV, almost
3 times the vacuum value. Close to the expected QCD phase the
$\rho$-resonance has melted, $\Gamma_\rho(T_c)$$\simeq$$m_\rho$.
``Rhosobar'' excitations ($\rho$$\to$$BN^{-1}$) lead to low-mass
strength in the $\rho$ spectral function that cannot be represented in
Breit-Wigner form.

Medium effects on the $\omega$ and $\phi$ have thus far received less
attention, especially in the context of ultrarelativistic heavy-ion
collisions (URHICs). For the $\omega$ we will employ the same approach
as for the $\rho$~\cite{Rapp:2000pe}. The predicted average $\omega$
width in the hadronic phase of URHICs is $\Gamma_\omega^{\rm
  med}$$\simeq$100~MeV~\cite{Rapp:2002mm}. For the $\phi$, collision
rates in a meson gas amount to a broadening by $\sim$20~MeV at
$T$=150~MeV~\cite{Alvarez-Ruso:2002ib}.  The dressing of the kaon cloud
is presumably the main effect for $\phi$ modifications in nuclear
matter, increasing its width by $\sim$25~MeV at saturation density
$\varrho_0$=0.16~fm$^{-3}$~\cite{Cabrera:2003wb}. Recent data on $\phi$
absorption in nuclear photoproduction suggest even larger
values~\cite{Ahn:2004id}. Since a comprehensive treatment of the $\phi$
in hot and dense matter is not available at present, we will consider
the $\phi$ width as a parameter.

\begin{figure}[!t]
\centerline{\includegraphics[width=0.45\textwidth]{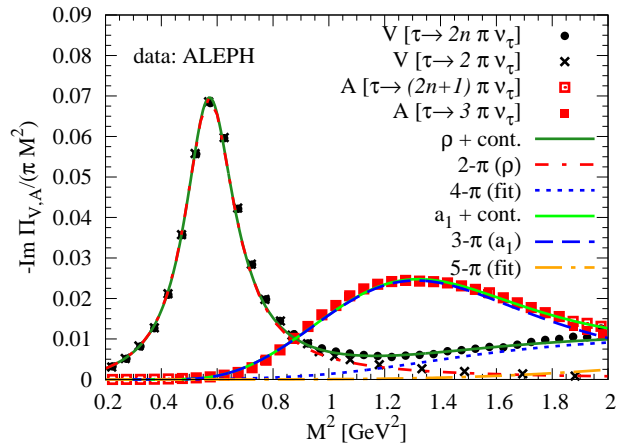}}
\caption{Free isovector-vector and -axialvector spectral functions 
as measured in hadronic $\tau$-decays~\cite{aleph98}, compared to
fits to 3- and 4-$\pi$ contributions (the 2-$\pi$ piece follows from 
the $\rho$).}
\label{fig_VAvac}
\end{figure}

Another important ingredient in our analysis are dilepton production
channels beyond the $\rho$, $\omega$ and $\phi$. For the free emission
rate these can be inferred from the inverse process of $e^+e^-$
annihilation, or hadronic $\tau$ decays as  
measured in $Z^0$ decays~\cite{aleph98}, enabling a 
decomposition of the (isovector
part of) e.m.~spectral function into 2- and 4-pion pieces. While the
former are saturated by the $\rho$-meson, we fit the latter by an 
appropriate (onset of a) continuum~\cite{Kapusta:1993hq}, 
cf.~Fig.~\ref{fig_VAvac}.
Rather than evaluating medium effects in the intermediate-mass
region (IMR; 1~GeV$\le$$M$$\le$3~GeV) in a phenomenological 
approach~\cite{Li:1998ma}, we here employ model-independent 
predictions based on chiral
symmetry. To leading order in temperature one finds a mixing of the 
free $V$ and $A$ correlators~\cite{Dey:1990ba},
\begin{equation}
\Pi_{V,A}(q) = (1-\varepsilon)~\Pi_{V,A}^{\rm vac}(q) +
\varepsilon~\Pi_{A,V}^{\rm vac}(q) \ ,
\label{mix}
\end{equation}
where $\varepsilon$=$2I(T)/f_\pi^2$ ($f_\pi$=93~MeV: pion decay
constant) is determined by the pion tadpole diagram with a loop integral
$I(T)$=$\int (d^3k/(2\pi)^3)~f^\pi(\omega_k;T)/\omega_k$ ($\omega_k$: pion
energy). For $\varepsilon$=1/2 (at $T_c^{\rm mix}$$\simeq$160~MeV) $V$
and $A$ correlators degenerate signaling chiral symmetry restoration.
Eq.~(\ref{mix}) holds in the soft pion limit, i.e., when 
neglecting the pion 4-momentum in $\Pi_{V,A}$. A more elaborate 
treatment~\cite{Steele:1997tv,Urban:2001uv} will broaden and reduce 
the enhancement around $M$$\simeq$1.1~GeV.  To estimate 
the mixing effect on the 4-$\pi$ contribution to dilepton emission, 
we evaluate the vector correlator, Eq.~(\ref{mix}), with a mixing 
parameter $\varepsilon=I(T,\mu_\pi)/2I(T_c,0)$, where critical temperature,
$T_c$=175~MeV, and pion chemical potentials ($\mu_\pi$$>$0) are in
accord with the thermal fireball evolution discussed below
($m_\pi$=139.6~MeV). The 2-pion piece has been removed as it is included
via the $\rho$.
 
Finally, we account for emission from the QGP, even though its
contribution at SPS energies is small. We employ the hard-thermal-loop
improved emission rate for $q\bar q$ annihilation~\cite{Braaten:1990wp},
which has the conceptually attractive feature that it closely coincides
with the rate in hadronic matter when both are extrapolated to the
expected phase transition region. This is suggestive for a kind of
quark-hadron duality~\cite{Rapp:1999us} and has the additional benefit
that the emission from the expanding fireball becomes rather insensitive
to details of how the phase transition is implemented (such as critical
temperature or ``latent heat'').

\textit{Thermal Fireball Model.}  To calculate dilepton spectra we need
to specify a space-time evolution of A-A collisions. Based on evidence
from hadronic spectra and abundances that the produced medium at SPS
energies reaches equilibrium~\cite{Braun-Munzinger:2003zd}, we adopt a
thermal fireball model focusing on central In-In with collective
expansion and hadrochemistry as estimated from data in heavier systems.
We use a cylindrical volume expansion~\cite{Rapp:1999us},
\begin{equation}
V_{\rm FB}(t) = (z_0+v_z t)~\pi~(r_\perp+0.5a_\perp t^2)^2     \ ,
\end{equation}
with transverse acceleration $a_\perp$=0.08$c^2/\text{fm}$, longitudinal
velocity $v_z$=$c$, formation time $\tau_0$=1fm/$c$ and initial
transverse radius $r_\perp$=5.15fm.  With hadrochemi\-cal freezeout at
$(\mu_N^{\rm chem}, T^{\rm chem})$=(232,175)~MeV and a total fireball
entropy of $S$=2630 (using a hadron-reso\-nance gas equation of state),
we have $\d N_{\text{ch}}/\d y$$\simeq$195$\simeq$$N_{\text{part}}$. 
Isentropic expansion allows to convert the entropy density, 
$s(t)$=$S/V(t)$, into temperature and baryon density. The evolution 
starts in the QGP ($T(\tau_0)$=197~MeV), passes through a mixed phase 
at $T^{\rm chem}$=$T_c$ and terminates at thermal freezeout
($T_{\text{fo}}$$\simeq$120~MeV). 
In the hadronic phase, meson chemical potentials ($\mu_{\pi,K,\eta}$) 
are introduced to preserve the observed hadron 
ratios~\cite{Rapp:1999us}. When applied to dilepton production, the
largest uncertainty resides in the fireball lifetime (being
proportional to the dilepton yields), controlled by $a_\perp$
and $T_{\text{fo}}$, which can be better constrained once hadronic data for
In-In are available. We emphasize, however, that all contributions
to the dilepton spectrum (QGP, $\rho$, $\omega$, $\phi$ and 4-$\pi$) are
tied to the \emph{same} evolution thus fixing their relative
weights.

\textit{Systematic Comparison to NA60 Data.} Thermal $\mu^+\mu^-$ 
spectra for central In-In are computed by convoluting the emission rate, 
Eq.~(\ref{rate}), over the fireball evolution,  
\begin{equation}
\label{spect}
\frac{\d N_{ll}}{\d M}=\int\limits_0^{t_{\text{fo}}} \d t~V_{\text{FB}}(t) 
\int \frac{d^3q}{q_0}~\frac{\d N_{ll}}{\d^4 x \d^4 q}~z_\pi^n
\, \frac{ M}{\Delta y}
 \,  A(M,q_t,y),
\end{equation}
where $A$ denotes the detector acceptance which has been carefully
tuned to NA60 simulations~\cite{na60acc}. The fugacity factor,
$z_\pi^n$=${\rm e}^{n\mu_\pi/T}$, accounts for chemical off-equilibrium
in the hadronic phase with $n$=2,3,4 for the $\rho$, $\omega$ and
4-$\pi$ contributions, respectively (for the mixing term in 
Eq.~(\ref{mix}) we adopt $\varepsilon~\Pi^{\rm vac}_A\propto 
z_\pi^4$~\cite{Urban:2001uv}).

\begin{figure}[!tbh]
\centerline{\includegraphics[width=0.45\textwidth]{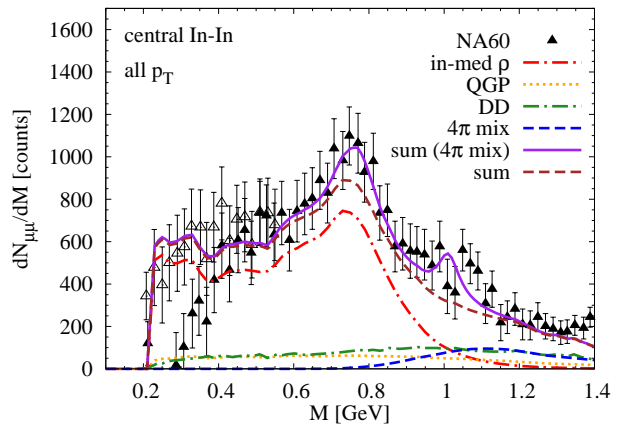}}
\caption{NA60 excess dimuons~\cite{Damjanovic:2005ni} in central
  In-In collisions at SPS compared to thermal dimuon radiation using
  in-medium e.m.~rates. The individual contributions arise from in-medium 
  $\rho$-mesons~\cite{Rapp:1999us} (dash-dotted red line), 4-$\pi$
  annihilation with chiral $V$-$A$ mixing (dashed blue line), QGP
  (dotted orange line) and correlated open charm (dash-dotted green
  line); the upper dashed line is the sum of the above, while the
  solid purple line additionally includes in-medium $\omega$ and $\phi$
  decays~\cite{Rapp:2000pe}.}
\label{fig_med}
\end{figure}

Earlier comparisons~\cite{Damjanovic:2005ni} of NA60 data to theoretical
predictions~\cite{Rapp:2004zh} have focused on the contribution from the
$\rho$-meson which dominates in the LMR. While the shape of the in-medium 
$\rho$ spectral function describes the experimental spectra well, the
absolute yields were overestimated by $\sim$20\%. This discrepancy can
be resolved by an increase of the transverse fireball expansion
($a_\perp$) which, on the one hand, reduces the fireball lifetime by
30\% (from 10 to 7~fm/$c$) and, on the other hand, generates harder
transverse-momentum ($q_t$) spectra, which is also welcome by
preliminary data~\cite{Damjanovic:2005ni}.  Consequently, the $\rho$
contribution is reduced, and once QGP emission and correlated
charm-decays ($D,\bar D \to \mu^+, \mu^- X$) are added, the spectra in
the LMR are very well described, cf.~Fig.~\ref{fig_med}.
Despite the strong $\rho$ broadening, the $\rho$+QGP+charm sources are
insufficient to account for the enhancement at $M$$\ge$0.9~GeV. This is
not surprising, as 4-$\pi$ contributions are expected to take over
(augmented by a pion fugacity factor, ${\rm e}^{4\mu_\pi/T}$).  Adding
the 4-$\pi$ piece with chiral mixing, Eq.~(\ref{mix}), nicely accounts
for the missing yield in the IMR, leading to a satisfactory overall
description (upper dashed line in Fig.~\ref{fig_med}).  Going one step
further still, we argue that in-medium decays of the narrow vector
mesons, $\omega$ and $\phi$, should be included.  Whereas their decays
after freezeout are subtracted as part of the ``cocktail'' assuming a
vacuum line shape, contributions whose width goes beyond the
experimental mass resolution will survive in the excess spectrum. With
the predicted in-medium $\omega$ spectral function~\cite{Rapp:2000pe}
the agreement between theory and data for $M$=0.7-0.8~GeV seems to
improve. The $\phi$ contribution is implemented with a two-kaon
fugacity, ${\rm e}^{2\mu_K/T}$, and a strangeness suppression factor
$\gamma_s^2$ ($\gamma_s$$\simeq$0.75 at SPS~\cite{Becattini:2003wp}).
The $\phi$ yield following from the fireball model is not incompatible
with data, but conclusions on the spectral shape cannot be drawn at
present.

\begin{figure}[!tbh]
\centerline{\includegraphics[width=0.45\textwidth]{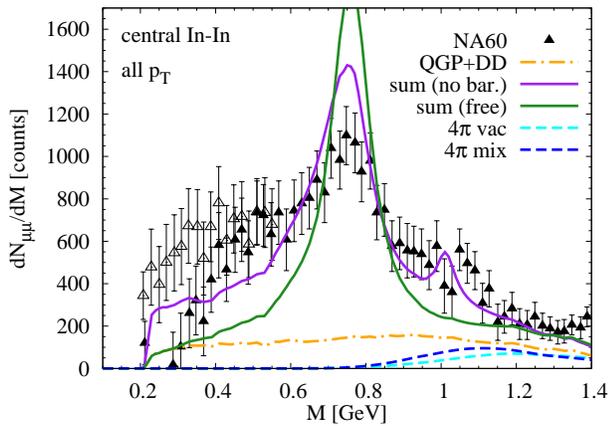}}
\caption{NA60 data~\cite{Damjanovic:2005ni} compared to thermal dimuon
  spectra using (i) in-medium $\rho$-, $\omega$- and $\phi$-mesons
  without baryon effects (+QGP+charm+in-med.-4-$\pi$; solid purple
  line), and (ii) free $\rho$ (+QGP+charm+free 4-$\pi$; solid green
  line).  }
\label{fig_vac}
\end{figure}
\begin{figure}[!t]
\centerline{\includegraphics[width=0.45\textwidth]{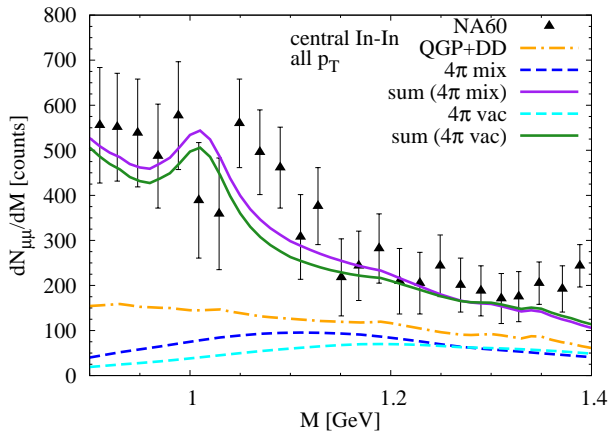}}
\caption{NA60 data~\cite{Damjanovic:2005ni} in the IMR compared to
  thermal dimuon spectra with different implementations of the 4-$\pi$
  contribution, using either its vacuum form (lower dashed line) or
  including chiral mixing (upper dashed line), and corresponding total
  spectra (lower and upper solid line, respectively).  }
\label{fig_mix}
\end{figure}

To better appreciate the relevance of the in-medium effects,
Fig.~\ref{fig_vac} shows two scenarios where (i) baryon-induced
interactions are neglected, and (ii) the vacuum e.m.~rate is employed.
The latter is ruled out; the meson-gas scenario, which differs from the
full result in Fig.~\ref{fig_med} by factors of 1.5-2 in the LMR, is not
favored by experiment either. Definite conclusions on the in-medium
$\rho$, especially on the baryon-driven enhancement close to the dimuon
threshold, require a reduction in systematic uncertainty, indicated by
the open and filled data points. Finally, Fig.~\ref{fig_mix} illustrates
the signature of chiral $V$-$A$ mixing in the IMR, where the full result
of Fig.~\ref{fig_med} is compared to a calculation with the 4-$\pi$
piece in the e.m.~rate replaced by its vacuum form (Eq.~(\ref{mix}) with
$\varepsilon$=0).  While the latter is not incompatible with the data,
an identification of medium effects requires better precision than
currently available (both from theory and experiment).

\textit{Conclusion.} We have conducted a quantitative investigation of
the dimuon excess measured by NA60 in In(158AGeV)-In. Focusing on
central collisions, where the notion of thermal radiation is most
adequate, we have shown that a medium modified e.m.~spectral function
properly accounts for absolute yields and spectral shape of the data.
While the overall normalization of the spectrum is subject to
uncertainties in the underlying fireball model (especially its
lifetime), the relative strength of the different components in the
spectrum ($\rho$, $\omega$ and $\phi$ decays, 4-$\pi$ type annihilation)
is fixed. Our results confirm the prevalent role of a strongly broadened
$\rho$-meson in the LMR, but also suggest substantial medium effects on
the line shapes of $\omega$ and $\phi$.  In addition, the IMR might bear
footprints of chiral $V$-$A$ mixing, which would support the notion that
the matter produced in A-A collisions at the SPS is close to chiral
restoration. A more quantitative connection to chiral order parameters,
e.g., by evaluating in-medium chiral sum rules~\cite{Kapusta:1993hq},
should be pursued with high priority.

\textit{Acknowledgments.} We are grateful to S.~Damjanovic and
H.J.~Specht for discussion and information on the NA60 acceptance.  HvH
thanks the A.~v.~Humboldt foundation for support via a Feodor Lynen
fellowship. This work was supported in part by a U.S. National Science
Foundation CAREER award under grant PHY-0449489.

\begin{flushleft}

\end{flushleft}


\end{document}